\documentclass{edp-conf}
\usepackage{graphicx}
\begin{document}
\TitreGlobal{SF2A 2002}

\title{Polar ring galaxies: formation and properties} 

\author{Bournaud, F.}\address{Ecole Normale Sup\'erieure, 45 rue d'Ulm, 75005 Paris, France}
\author{Combes, F.}\address{LERMA, Observatoire de Paris, 61 av. de l'Observatoire, 75014 Paris, France}

\runningtitle{Polar ring galaxies}

\setcounter{page}{237}

\index{Bournaud, F.}
\index{Combes, F.}

\maketitle

\begin{abstract}
Formation scenarios for polar ring galaxies are studied through N-body simulations that are compared with existing observations. It is shown that polar rings are likely to be formed by tidal accretion of the polar material from a gas rich donor galaxy. The distribution of dark matter in polar ring galaxies is studied: dark halos seem to be flattened towards the polar rings.
\end{abstract}

\section{Introduction}
Polar ring galaxies are peculiar systems that show two nearly orthogonal components. Two formation scenarios for polar rings have been proposed: galaxy mergers (Bekki, 1998) and tidal accretion (Reshetnikov \& Sotnikova, 1997). They are studied through numerical simulations including gas dynamics, star formation and stellar mass-loss, and compared to observations.

\section{Formation scenarios}\label{sect1}

\begin{figure}[h]
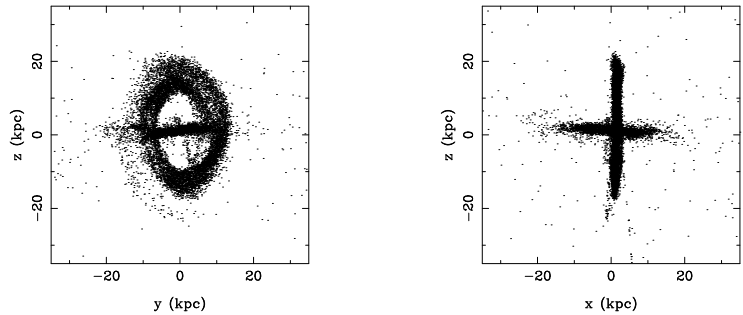

   \centering
   \includegraphics[angle=270,width=4cm]{bournaud_fig1.ps}
   \hspace{1.5cm}
   \includegraphics[angle=270,width=4cm]{bournaud_fig2.ps}
      \caption{Example of polar ring formed by tidal accretion from a gas rich-donor galaxy in a numerical simulation. The dark halo is here spherical, but the flat host galaxy makes the polar ring be significantly eccentric.}
       \label{fig1}
   \end{figure}

Numerical studies of formation scenarios (Fig.~\ref{fig1}) show that both scenarios can reproduce the observations with a certain range of parameters. Polar rings are also stable in both scenarios. However, constraints on physical parameters are stronger for the merging scenario, which reduces its probability, while the formation of polar ring by gas accretion is much more probable (Bournaud \& Combes 2002).

Numerical simulations show that according to the merging scenario, the polar ring galaxy is embedded in a faint stellar halo, which is not the case in the accretion scenario. We show from observations that such a stellar halo is not detected. A luminosity model is build with two components (the host galaxy and the ring) and no third component (the stellar halo) is required to account for observations. This implies that in the merging scenario, one of the galaxy invoked in the merger should have contained more then 60\% of gas for a total mass of $2\cdot10^{10}$ M$_\odot$, which is a strong additional constraint (only satisfied for gas-rich LSB). 

In the polar ring galaxy NGC~660, the ring is inclined of 45 degrees with respect to the polar axis, and the host galaxy is gas-rich. Both properties cannot be explained only by the accretion scenario.

\section{Dark matter in polar ring galaxies}\label{sect2}
Observations of polar ring galaxies are compared to the Tully-Fisher relation for spiral galaxies (Iodice et al. 2002). For a given luminosity, polar ring galaxies show abnormally large velocities. The luminosity, in optical bands, is related to the bright host galaxy, while velocities, measured in HI, are related to the gas-rich polar ring. The velocities in the polar ring are then larger than expected from the mass of the host galaxy. Numerical models, that include the polar ring self-gravity and the dark halo shape, are used to interprete these observations. The eccentricity of polar rings (Fig.~\ref{fig1}) is a fundamental parameter when one studies their kinematics in order to derive the shape of their dark halo. We show that the observations of large velocities in polar rings are accounted for only when the dark halo is flattened along the polar ring. According to simulations, such halo shapes would be explained by the accretion scenario, provided that some dark matter is cold molecular gas, accreted in the same time as the polar ring.

\end{document}